\documentclass[11pt,a4paper]{article}
\pdfoutput=1 

\usepackage{jheppub} 

\newcommand{\mre}{\mathrm{e}}
\newcommand{\mrd}{\mathrm{d}}

\newcommand{\Z}{{\mathbb{Z}}}

\newcommand{\one}{{\bf 1}}

\newcommand{\mcA}{\mathcal{A}}

\newcommand{\mcL}{\mathcal{L}} 
\newcommand{\mcW}{\mathcal{W}}
\newcommand{\mcWov}{\overline{\mathcal{W}}}
\newcommand{\cov}{\overline{c}}
\newcommand{\Cov}{\overline{C}}

\newcommand{\IDR}{\overline{I}}
\newcommand{\Wov}{\overline{W}}
\newcommand{\Uov}{\overline{U}}

\newcommand{\psump}{\sum_{p}\rule{0pt}{2.5ex}'\;}

\newcommand{\order}[1]{\mathcal{O}\left(#1\right)}

\newcommand{\bfp}{\mathbf{p}}
\newcommand{\bfx}{\mathbf{x}}
\newcommand{\bfy}{\mathbf{y}}
\newcommand{\bfz}{\mathbf{z}}

\newcommand{\Nf}{N_\mathrm{f}}

\newcommand{\Ls}{L_s}
\newcommand{\Lt}{L_t}
\newcommand{\Lhat}{\widehat{L}}
\newcommand{\ellhat}{\hat{\ell}}
\newcommand{\ds}{d_s}

\newcommand{\msbar}{{\rm \overline{MS\kern-0.14em}\kern0.14em}}

\newcommand{\Ns}{N_1}

\newcommand{\VdD}{\overline{V}_D}

\newcommand{\mcLov}{\overline{\mathcal{L}}}

\DeclareMathOperator{\tr}{tr}

\newcommand{\Lr}[1]{L_{#1}^\mathrm{r}}
\newcommand{\lr}[1]{l_{#1}^\mathrm{r}}

\newcommand{\SUN}{\mathrm{SU}(N)}
\newcommand{\SUtw}{\mathrm{SU}(2)}

\newcommand{\Of}{\mathrm{O}(4)}
\newcommand{\On}{\mathrm{O}(n)}
\newcommand{\SSUN}{\mathrm{SU}(N)\times\mathrm{SU}(N)}
\newcommand{\SSUtw}{\mathrm{SU}(2)\times\mathrm{SU}(2)}
\newcommand{\SSUth}{\mathrm{SU}(3)\times\mathrm{SU}(3)}

\title{\boldmath Finite volume mass gap and free energy of the $\SSUN$
chiral sigma model} 

\author[a]{F.\ Niedermayer} \author[b]{and P.\ Weisz}

\affiliation[a]{Albert Einstein Center for Fundamental Physics, \\
  Institute for Theoretical Physics, University of Bern, 
  Switzerland} 
\affiliation[b]{Max-Planck-Institut f\"ur Physik, 80805
  Munich, Germany}

\emailAdd{niedermayer@itp.unibe.ch} \emailAdd{pew@mpp.mpg.de}

\abstract{We compute the free energy in the presence of a chemical potential
  coupled to a conserved charge in the effective $\SSUN$ 
  scalar field theory to third order for asymmetric
  volumes in general $d$--dimensions, using dimensional regularization (DR).
  We also compute the mass gap in a finite box with periodic boundary
  conditions.} 

\subheader{\hfill MPP-2017-63}

\begin{document}

\maketitle

\section{Introduction}
\label{Introduction}

Chiral perturbation theory 
($\chi$PT)~\cite{Weinberg:1978kz,Gasser:1984gg}
is the effective theory describing the low energy dynamics of 
the lowest lying pseudoscalar mesons. The parameters of the
theory are couplings appearing in the effective chiral
Lagrangian, the pion decay constant $F_\pi$ 
($=F$ in the chiral limit) and other low energy constants (LEC's). 
These parameters can be determined by phenomenology, 
or by lattice simulations of QCD. For a detailed summary of 
various determinations of the LEC's the reader is 
referred to the FLAG review~\cite{Aoki:2016frl}.

For $\Nf=2$ the relevant $\chi$PT has $\SSUtw\simeq\Of$ symmetry. 
As a consequence in the past many theoretical $\chi$PT computations, 
in particular those pertaining to finite volume,
have been performed for the slightly simpler model with $\On$ symmetry.
One special environment is the so called
$\delta$--regime first discussed by Leutwyler~\cite{Leutwyler:1987ak}
where the system is in a periodic spatial box of sides $\Ls$
and $m_\pi\Ls$ is small (i.e.\ small or zero quark mass) 
whereas $F_\pi\Ls$ is large.
In 2009 Hasenfratz~\cite{Hasenfratz:2009mp} computed the mass 
gap in the delta-regime to third order $\chi$PT with the hope 
that a comparison with a precise lattice measurement of the low-lying 
stable masses in this regime may be used to determine some combination 
of the LEC's.

In a previous paper~\cite{Niedermayer:2016yll} we computed 
the change in the free energy due to a chemical potential
coupled to a conserved charge in the non-linear O($n$) sigma model with two
regularizations, lattice regularization (with standard action) and
DR in a general $d$-dimensional asymmetric
volume with periodic boundary conditions (pbc) in all directions. 
This freedom allowed us for $d=4$ to establish two
independent relations among the 4-derivative couplings appearing in the
effective Langrangians and in turn this allows
conversion of results for physical quantities computed by the lattice
regularization to those involving scales introduced in DR.  

In particular we could convert the computation of the mass gap
in a periodic box, by Niedermayer and Weiermann~\cite{Niedermayer:2010mx} 
using lattice regularization to a result involving parameters of the
dimensionally regularized effective theory, and we verified this
result by a direct computation~\cite{Niedermayer:2016yll} (which 
disagrees slightly with the previous computation~\cite{Hasenfratz:2009mp}). 

Although $\Nf=2$ is the phenomenologically most relevant case
due to the low mass of the physical pions, $\chi$PT with
$\Nf>2$ can also have useful applications~\cite{Aoki:2016frl}.
With this in mind in this paper we extend the computations 
to the case of $\SSUN$. After recollecting the 
structure of the effective Lagrangian in the next section
we compute the free energy is in sect.~\ref{chempotcont} 
and the mass gap in a finite periodic box in sect.~\ref{massgap}. 

In this paper we do not analyze explicit chiral symmetry breaking. 
In QCD the effect of including a small quark mass
on the finite volume spectrum has been computed for $\Nf=2$ 
to leading order in~\cite{Leutwyler:1987ak}, and to next-to-leading order 
by Weingart~\cite{Weingart:2010yv,Weingart:2010zz}. 
Furthermore Matzelle and Tiburzi~\cite{Matzelle:2015lqk} have
studied the effect of small symmetry breaking in the 
quantum mechanical (QM) rotator picture
($\Nf=2$), and extended the results to small non-zero temperatures.
In a related recent paper~\cite{Niedermayer:2017uyr} 
we have computed the isospin susceptibility
in the effective O($n$) scalar field theory, to third order $\chi$PT
in the delta-regime using the QM rotator picture
including an explicit symmetry breaking term, and showed
consistency with standard $\chi$PT computations.

\section{The effective Lagrangian}

The dynamical fields are matrices $U(x)\in\SUN$.
In the chiral limit the action is invariant under
global $\SUN_L\times\SUN_R$ transformations of the fields 
\begin{equation}
  U(x)\rightarrow g_L U(x) g_R^{\dagger}\,.
\end{equation}
In this limit the leading order effective Lagrangian 
is given by~\cite{Gasser:1984gg}:
\begin{equation}
  \mcL_1 = \frac{1}{4g_0^2}
  \tr\left( \partial_\mu U^\dagger \partial_\mu U \right)\,.
\end{equation}
For $N\ge 4$ there are four linearly 
independent\footnote{up to higher derivatives}
four-derivative terms in the effective Lagrangian~\cite{Gasser:1984gg} 
\begin{equation}
  \mcL_2 = \sum_{i=0}^3\frac{G_4^{(i)}}{4}\mcL_2^{(i)}\,,
  \label{mcL2}
\end{equation}
with
\begin{align}
  \mcL_2^{(0)} & = \tr\left( \partial_\mu U^\dagger \partial_\nu U
    \partial_\mu U^\dagger \partial_\nu U \right)\,,
  \label{L20}
  \\
  \mcL_2^{(1)} & = \tr^2 \left( \partial_\mu U^\dagger \partial_\mu U \right)\,,
  \label{L21}
  \\
  \mcL_2^{(2)} & = \tr\left( \partial_\mu U^\dagger \partial_\nu U \right)
  \tr\left( \partial_\mu U^\dagger \partial_\nu U \right)\,,
  \label{L22}
  \\
  \mcL_2^{(3)} & = \tr\left( \partial_\mu U^\dagger \partial_\mu U
    \partial_\nu U^\dagger \partial_\nu U \right)\,.
  \label{L23}
\end{align}
The 4-derivative couplings in~\eqref{mcL2} are related to the standard 
ones~\cite{Gasser:1984gg} as $G_4^{(i)}=-4 L_i$. Since we work here 
in Euclidean space-time, our couplings differ in sign.\footnote{To avoid
confusion with the box size $L_\mu$ we shall use the renormalized couplings 
$\Lr{i}$ only in the final results.}
Note also the absence of the 4-derivative term 
$\tr(\Box U^\dagger \Box U)$ in the above list;
As explained in~\cite{Leutwyler:1991mz},
this term can be eliminated by redefinition of the field $U$.
The argument is reproduced for completeness in Appendix~\ref{AppC}.

For $N<4$ these four operators are not all independent. 
One has~\cite{Leutwyler:1991mz} 
\begin{align}
  \mcL_2^{(0)}&=-\frac12 \mcL_2^{(1)}+\mcL_2^{(2)} \,, 
  \qquad\qquad  \mcL_2^{(3)}=\frac12 \mcL_2^{(1)} \,, & (N&=2) \,,
  \label{mcLSU2id} \\
  \mcL_2^{(0)}&=\frac12 \mcL_2^{(1)} +\mcL_2^{(2)} - 2\mcL_2^{(3)} \,,  & (N&=3) \,. 
  \label{mcLSU3id}
\end{align}
A proof of~\eqref{mcLSU3id} is given in Appendix~\ref{AppB}.
Accordingly, in~\eqref{mcL2} one can restrict the summation to $i=1,2$ 
for $N=2$ and to $i=1,2,3$ for $N=3$.

From these relations it follows that the results obtained for general $N$
should at $N=2$ be invariant under the transformation
\begin{equation}
  \begin{aligned}
    G_4^{(0)}   &\to G_4^{(0)}+\alpha_1 \,,
    & G_4^{(1)} &\to G_4^{(1)}+\alpha_2 \,, \\
    G_4^{(2)}   &\to G_4^{(2)}-\alpha_1 \,,
    & G_4^{(3)} &\to G_4^{(3)}+\alpha_1-2\alpha_2\,, \quad (N=2) \,,
  \end{aligned}
  \label{GtransSU2}
\end{equation}
while at $N=3$ under
\begin{equation}
  \begin{aligned}
    G_4^{(0)}  &\to G_4^{(0)} + 2\alpha \,,
    &G_4^{(1)} &\to G_4^{(1)} - \alpha \,, \\
    G_4^{(2)}  &\to G_4^{(2)} - 2\alpha \,,
    &G_4^{(3)} &\to G_4^{(3)} + 4\alpha\,, \quad (N=3) \,.
  \end{aligned}
  \label{GtransSU3}
\end{equation}

The $\SSUN$ model for $N=2$ flavors
is equivalent to the $\Of$ non-linear sigma model~\cite{Gasser:1983yg}
(with fields $S_i\,,\,\,i=0,\dots,3$ and $S^2=1$) where
\begin{equation}
  \mcLov_1 = \frac{1}{2g_0^2}\left(\partial_\mu S\cdot\partial_\mu S\right)\,,
\end{equation}
and
\begin{equation}
  \mcLov_2 = \sum_{i=2,3}\frac{g_4^{(i)}}{4}\mcLov_2^{(i)}
\end{equation}
with
\begin{align}
  \mcLov_2^{(2)}& = \left(\partial_\mu S \cdot \partial_\mu S  \right)^2\,,
  \\
  \mcLov_2^{(3)}& =\left(\partial_\mu S \cdot \partial_\nu S \right)
  \left(\partial_\mu S \cdot \partial_\nu S \right)\,.
\end{align}

Writing
\begin{equation}
  U(x) = S_0(x) + i \sum_{a=1}^3 \sigma^a S_a(x)\,, 
\qquad S^2(x) = 1\,, 
\end{equation}
where $\sigma^a$ are the Pauli matrices, one obtains
\begin{equation}
  \mcLov_2^{(2)} = \frac14 \mcL_2^{(1)} \,, \qquad  
  \mcLov_2^{(3)} = \frac14 \mcL_2^{(2)}\,. 
  \label{SU2O4}
\end{equation}
This leads to the identification~\cite{Gasser:1984gg} 
\begin{equation}
  g_4^{(2)} = 4G_4^{(1)} \,, \qquad g_4^{(3)} = 4G_4^{(2)} \,.
  \label{GLid}
\end{equation}
These and the relations~\eqref{GtransSU2},~\eqref{GtransSU3} can serve as
checks on the final results.

\subsection{Perturbative expansion}

Here we work in a continuum volume $V=\Lt\times\Ls^{\ds}\,,
$\,\,\,$\ds=d-1$. In this section we impose periodic boundary
conditions (pbc) on the dynamical variables in all directions.  
We dimensionally regularize by adding $q$
extra compact dimensions of size $\Lhat$ (also with pbc) and analytically
continue the resulting loop formulae to $q=-2\epsilon$. We define
$D=d+q\,,V_D=V\Lhat^q$\,, and the aspect ratios  
$\ell=\Lt/\Ls\,,\ellhat\equiv\Lhat/\Ls$\footnote{It is advantageous to 
treat these extra dimensions with a different size, since an extra check 
of the calculation is provided by the requirement that
physical quantities are independent of this choice.}.

For the perturbative expansion we parameterize $U$ with scalar fields 
$\xi_a(x)$\footnote{For SU(2) the identification to the O(4) fields 
$S_a$, $a=1,2,3$ is $S_a=g_0\pi_a=\xi_a\sin(g_0|\xi|)/|\xi|$,
where
$|\xi|=\sqrt{\sum_b \xi_b\xi_b}$.}
\begin{equation}\label{Uparam}
  U(x)=u\Uov(x)\,,\,\,\,\Uov(x)=\exp\left(ig_0\xi(x)\right)\,,
\end{equation}
where $u$ is a constant matrix and 
\begin{equation}
  \xi=\sum_{a=1}^{\Ns}\lambda^a\xi_a\,,
\end{equation}
where the hermitian $\lambda$-matrices are defined 
and some of their properties noted in Appendix~\ref{AppA}.
Further
\begin{equation}
  \Ns\equiv N^2-1\,,
\end{equation}
and the fields $\xi$ satisfy the constraints
\begin{equation}
  \int_x \xi_a(x)=0\,,\,\,\,\,\,\,\forall a\,.
\end{equation}

$A_{2,\mathrm{eff}}$ has a perturbative expansion
\begin{equation}
  A_{2,\mathrm{eff}} = A_{2,0} + g_0^2 A_{2,1} + g_0^4 A_{2,2} 
  +\order{g_0^6}\,,
  \label{Aeff_cont} 
\end{equation}
where
\begin{align}
  A_{2,0} & = \frac12\int_x\partial_\mu\xi_a(x)\partial_\mu\xi_a(x)\,,
  \label{A20_cont}
  \\
  A_{2,1} & = A_{2,1}^{(a)}+A_{2,1}^{(b)}\,, \label{A21_cont}
  \\
  A_{2,1}^{(a)} & = \frac{N}{3V_D}\int_x\sum_a\xi_a(x)\xi_a(x)\,,
 \label{A21a_cont}
  \\
  A_{2,1}^{(b)} & = \frac{1}{48}\int_x\tr\left(
  \left[\xi(x),\partial_\mu\xi(x)\right]^2\right)\,,
 \label{A21b_cont}
\end{align}
and
\begin{align}
  A_{2,2} & = A_{2,2}^{(a)}+A_{2,2}^{(b)}+A_{2,2}^{(c)}\,, \label{A22_cont}
  \\
  A_{2,2}^{(a)} & =\frac{1}{1440V_D}\int_x
  \sum_a\tr\left(\lambda^a\left[\xi(x),\left[\xi(x),\left[\xi(x),
  \left[\xi(x),\lambda^a\right]\right]\right]\right]\right)\,,
 \\
  A_{2,2}^{(b)} & =\frac{1}{1152V_D^2}\int_{x y}\sum_{a,b}
\tr\left(\lambda^b\left[\xi(x),\left[\xi(x),\lambda^a\right]\right]\right)
\tr\left(\lambda^a\left[\xi(y),\left[\xi(y),\lambda^b\right]\right]\right)\,,
  \\
  A_{2,2}^{(c)} & =\frac{1}{1440}\int_x\tr\left(
  \left[\xi(x),\left[\xi(x),\partial_\mu\xi(x)\right]\right]^2\right)\,,
\end{align}
where the terms $A_{2,1}^{(a)},A_{2,2}^{(a)},A_{2,2}^{(b)}$
come from the zero mode action derived in Appendix~\ref{AppD}.

The total effective action has a perturbative expansion of the form
\begin{equation}
  \mcA=\sum_{r=0}\mcA_r g_0^{2r}\,, 
\end{equation}
with
\begin{equation}
  \mcA_r=A_{2,r}+\sum_{i=0}^3\frac{G_4^{(i)}}{4}A_{4,r}^{(i)}\,.
\end{equation}
Note
\begin{equation}
  A_{4,0}^{(i)}=0=A_{4,1}^{(i)}\,,\,\,\,\forall i\,,
\end{equation}
and
\begin{align}
  A_{4,2}^{(0)} & = \int_x\tr\left(\partial_\mu \xi(x)\partial_\nu \xi(x)
    \partial_\mu \xi(x)\partial_\nu\xi(x) \right)\,,
  \\
  A_{4,2}^{(1)} & = \int_x\tr^2\left(\partial_\mu\xi(x)\partial_\mu\xi(x)
  \right)\,,
  \\
  A_{4,2}^{(2)} & = \int_x\tr\left(\partial_\mu\xi(x) \partial_\nu\xi(x) \right)
  \tr\left( \partial_\mu\xi(x) \partial_\nu\xi(x) \right)\,,
  \\
  A_{4,2}^{(3)} & = \int_x\tr\left( \partial_\mu\xi(x)\partial_\mu\xi(x)
    \partial_\nu\xi(x)\partial_\nu\xi(x) \right)\,.
\end{align}

The free 2-point function is given by
\begin{equation}
  \langle\xi_a(x)\xi_b(y)\rangle_0=\delta_{ab}G(x-y)\,,
\end{equation} 
with propagator
\begin{equation}
  G(x)=\frac{1}{V_D}\psump \frac{\mre^{ipx}}{p^2}\,,
\end{equation}
where the sum is over momenta $p_\mu=2\pi n_\mu/L_\mu\,,\,\,n_\mu\in\Z$ and
the prime on the sum means that $p=0$ is omitted.

\section{The chemical potential}
\label{chempotcont}

The chemical potential $h$ is introduced by the substitution:
\begin{equation}
  \partial_0 \to \partial_0 + h\left[\frac{\lambda^3}{2},\cdot\right]\,.
\end{equation}

This gives an additional $h$-dependent part $\mcA_h$ to the total action of
the form
\begin{equation}
  \mcA_h=A_{2h}+\sum_{i=0}^3\frac{G_4^{(i)}}{4}A_{4h}^{(i)}\,.
\end{equation}
Further writing
\begin{align}
  &A_{2h}=ihB_2+h^2C_2+\dots\,,\\
  &A_{4h}^{(i)}=ihB_4^{(i)}+h^2 C_4^{(i)}+\dots
  \label{A4h_DR}\,,
\end{align}
we have
\begin{align}
  B_2 & =  -\frac{i}{4g_0^2}\int_x \tr
  \left(\lambda^3\left[U(x),\partial_0 U^\dagger(x)\right]\right)\,,
  \\ 
  C_2 & = \frac{1}{16g_0^2}\int_x 
  \tr \left(\left[\lambda^3,U(x)\right]
    \left[\lambda^3,U^\dagger(x)\right]\right)\,.
\end{align}

The 4-derivative operators $B_4^{(i)},C_4^{(i)}$ are given
in Appendix~\ref{AppF}.

The $h$-dependent part of the free energy $f_h$ is defined as 
\begin{equation}
  \mre^{- V f_h } = \langle \mre^{-\mcA_h}
  \rangle_\mcA\, = 1 - \langle \mcA_h \rangle_\mcA + \frac12 \langle \mcA_h^2
  \rangle_\mcA + \ldots
  \label{free_energy}
\end{equation}
giving up to the order $h^2$:
\begin{equation}
  V f_h  = \langle \mcA_h\rangle_\mcA-\frac12\langle \mcA_h^2\rangle_\mcA
  +\frac12 \langle \mcA_h\rangle_\mcA^2+\dots
\end{equation}
Note for an observable $X$:
\begin{equation}
  \langle X \rangle_\mcA = \langle X \rangle_0
  - g_0^2 \langle X \mcA_1\rangle_0^c
  - g_0^4 \langle X \mcA_2\rangle_0^c
    + \frac12 g_0^4 \langle X \mcA_1^2 \rangle_0^c + \ldots
\end{equation}
Now 
\begin{equation}
  \langle B_2\rangle_\mcA=0=\langle B_4^{(i)}\rangle_\mcA
  \,\,\,\,\,\,\forall i\,,
\end{equation}
so we have
\begin{equation}
  \chi =-2\sum_{s=1}^5F_s\,,
\end{equation}
with
\begin{align}
  F_1 & = \frac{1}{V_D}\langle C_2 \rangle_\mcA\,, \label{F1defcont}
  \\
  F_2  & = \frac12\frac{1}{V_D}\langle B_2^2\rangle_\mcA\,,  \label{F2defcont}
  \\
  F_3 & = \sum_{i=0}^3\frac{G_4^{(i)}}{4}\frac{1}{V_D} \langle
  C_4^{(i)}\rangle_\mcA\,,   \label{F3defcont}
  \\
  F_4 & =
  \sum_{i=0}^3\frac{G_4^{(i)}}{4}\frac{1}{V_D}\langle
  B_2B_4^{(i)}\rangle_\mcA\,,   \label{F4defcont}
  \\
  F_5 & = \frac12\sum_{i,j=0}^3\frac{G_4^{(i)}}{4}
  \frac{G_4^{(j)}}{4} \frac{1}{V_D}\langle
  B_4^{(i)}B_4^{(j)}\rangle_\mcA\,.  \label{F5defcont}
\end{align}

Averaging over the zero modes, 
denoting $\Uov(x)=\mre^{ig_0\lambda\xi(x)}$, 
\begin{equation}
  \begin{aligned}
    \frac{1}{V_D}\int\mrd u\,C_2&=\frac{1}{8g_0^2}\int_x\mrd u\, 
    \tr \left( 
      \lambda^3u\Uov(x)\lambda^3\Uov^\dagger(x) u^\dagger
      -\left(\lambda^3\right)^2\right)
    \\
    &=-\frac{1}{4g_0^2}\,,
  \end{aligned}
\end{equation}
where we used~\eqref{intuudagger}. So
\begin{equation}
  F_1=-\frac{1}{4g_0^2}\,.
\end{equation}

Next
\begin{equation}
 \frac{1}{V_D}\int\mrd u\,B_2^2 = \frac{1}{g_0^4}W\,,
 \label{Bsq_cont}
\end{equation}
with $W$ given by 
\begin{equation}
  W=-\frac{1}{16V_D}\int_{x y}\int\mrd u\,
  \tr\left(\lambda^3\left[u\Uov(x),
      \partial_0\Uov^\dagger(x)u^\dagger\right]\right)
  \tr\left(\lambda^3\left[u\Uov(y),
      \partial_0\Uov^\dagger(y)u^\dagger\right]\right)\,.
\end{equation}
For the averages we have
\begin{equation}
  \langle W \rangle_\mcA=\frac{1}{\Ns}\langle [W]\rangle_\mcA\,,
\end{equation}
where $[W]$ is obtained from $W$ by replacing
$\lambda^3_{ij}\lambda^3_{kl}$ by 
$\sum_a \lambda^a_{ij}\lambda^a_{kl}$. Using completeness
in the form~\eqref{compl1} we get
\begin{align}
  [ W ] &=-\frac{1}{8V_D}\int_{x y}\int\mrd u\,
  \tr\left(\left[u\Uov(x),\partial_0\Uov^\dagger(x)u^\dagger\right]
    \left[u\Uov(y),\partial_0\Uov^\dagger(y)u^\dagger\right]\right)
  \nonumber
  \\
  &=\frac{1}{8V_D}\int_{x y}\int\mrd u\,
  \tr\left(J_-(x)J_-(y)+J_+(x)J_+(y)\right.
  \nonumber
  \\
  &\quad\quad\quad\quad\quad\quad
  \left. +u J_-(x)u^\dagger J_+(y)+J_+(x)uJ_-(y)u^\dagger\right)\,,
\end{align}
where
\begin{equation}
  J_+(x)\equiv i\partial_0\Uov^\dagger(x)\Uov(x)\,,\,\,\,
  J_-(x)\equiv i\partial_0\Uov(x)\Uov^\dagger(x)\,.
\end{equation}
Note $J_\pm$ are hermitian $J_{\pm}^\dagger(x)=J_\pm(x)$ and traceless
\begin{equation} 
  \tr(J_\pm(x))=0\,,
\end{equation}
and have a perturbative expansion\footnote{For $a,b$ in the Lie algebra 
$\mathrm{Ad}(a)b=[a,b]$}
\begin{align}
  J_\pm(x)&=\pm g_0\left(\frac{\exp(\mathrm{Ad}[\pm ig_0\xi(x)])-1}
    {\mathrm{Ad}[\pm ig_0\xi(x)]}\right)\partial_0\xi(x)
  \\
  &=\pm g_0\sum_{r=1}^\infty\frac{1}{r!}
  \left(\mathrm{Ad}[\pm ig_0\xi(x)]\right)^{r-1}\partial_0\xi(x)
  \nonumber\\
  &=\pm g_0\partial_0\xi(x)
  +i\frac{g_0^2}{2}\left[\xi(x),\partial_0\xi(x)\right]
  \mp \frac{g_0^3}{6}\left[\xi(x),\left[\xi(x),\partial_0\xi(x)\right]\right]
  \nonumber\\
  &-i\frac{g_0^4}{24}
  \left[\xi(x),\left[\xi(x),\left[\xi(x),\partial_0\xi(x)\right]\right]\right]
  +\order{g_0^5}\,.
  \label{Jpert}
\end{align}
Note that for pbc
\begin{equation}
  \int_x J_\pm(x) = 0 \,.
\end{equation}

Using~\eqref{intuudagger} we get simply
\begin{equation}
  [W]=\frac{1}{8V_D}\int_{x y}\tr\left(J_-(x)J_-(y)+J_+(x)J_+(y)\right)\,.
\end{equation}
This has a perturbative expansion
\begin{equation}
  [W] = g_0^4 W_2 + g_0^6 W_3 + \ldots
  \label{Wa_cont}
\end{equation}
with
\begin{equation}
  \begin{aligned}
    W_2&=-\frac{1}{16V_D}\int_{x y}
    \tr\left(\left[\xi(x),\partial_0\xi(x)\right]
      \left[\xi(y),\partial_0\xi(y)\right]\right) 
    \\
    &=\frac{1}{2V_D}f_{abe}f_{cde}\int_{x y}\xi_a(x)\partial_0\xi_b(x)
    \xi_c(y)\partial_0\xi_d(y)\,,
  \end{aligned}
  \label{W2_cont}
\end{equation}
and
\begin{equation}
  W_3=W_3^{(1)}+W_3^{(2)}\,,
\end{equation}
with
\begin{align}
  & W_3^{(1)} = \frac{1}{144 V_D}\int_{xy}
  \tr\left(\left[\xi(x),\left[\xi(x),\partial_0\xi(x)\right]\right]
    \left[\xi(y),\left[\xi(y),\partial_0\xi(y)\right]\right]\right)\,,
  \\
  & W_3^{(2)} = \frac{1}{96 V_D}\int_{xy}
  \tr\left(\left[\xi(x),\partial_0\xi(x)\right]
    \left[\xi(y),\left[\xi(y),\left[\xi(y),\partial_0\xi(y)\right]\right]\right]
  \right)\,.
\end{align}

Expanding~\eqref{F2defcont} in a perturbative series
\begin{equation}
  F_2=\sum_{r=0}^\infty F_{2,r}g_0^{2r}\,,
\end{equation}
we have at leading order
\begin{equation}
  F_{2,0}=\frac{1}{2\Ns}\langle W_2\rangle_0\,.
  \label{F20}
\end{equation}
Now
\begin{equation}
  \begin{aligned}
    \langle W_2\rangle_0 &=\frac18 Z_1\int_x\left[\partial_0 G(x)\right]^2
    \\
    &=\frac18 Z_1 \IDR_{21}\,.
  \end{aligned}
\end{equation}
Here $Z_1$ is a group factor defined in~\eqref{Z1def} in Appendix~\ref{AppA}
where also other such factors $Z_i\,,i=2,\dots,8$ appearing below
are defined and evaluated.
Further the dimensionally regularized sums $\IDR_{nm}$ are formally 
defined by 
\begin{equation}
  \IDR_{nm} = \frac{1}{V_D} {\sum_p}'
  \;\frac{\left(p_0^2\right)^m}{\left(p^2\right)^n} \,.
  \label{IDRnm}
\end{equation}
So we have at leading order
\begin{equation}
    F_{2,0}=\frac{N}{2}\IDR_{21}\,.
  \label{F20cont}
\end{equation}

At next order
\begin{equation}
  F_{2,1}=\frac{1}{2\Ns}\left[\langle
    W_3\rangle_0 -\langle W_2A_{2,1}\rangle_0^c \right]\,.
  \label{F21cont}
\end{equation}
First\footnote{we used 
$\int_y\partial_0^y\left[G(x-y)^2\partial_0^{x} G(x-y)\right]=0$}
\begin{equation}
  \langle W_3^{(1)}\rangle_0 = \frac{1}{96}(Z_2+Z_3)\Wov
=\frac12  N^2\Ns\Wov\,, 
  \label{W31avx}
\end{equation}
where
\begin{equation}
  \Wov = -\int_x G(x)^2\partial_0^2 G(x)\,.
  \label{Wov}
\end{equation}
This 2-loop function, the ``massless sunset diagram'', is calculated
in detail in~\cite{Niedermayer:2016ilf}.

Secondly
\begin{align}
  \langle W_3^{(2)}\rangle_0 &= \frac{1}{48} Z_5G(0)\int_x
  \left[\partial_0 G(x)\right]^2
  \label{W32av}
\\
  &= -\frac53 N^2\Ns\IDR_{10}\IDR_{21}\,.
  \label{W32avx}
\end{align}

Next
\begin{equation}
  \begin{aligned}
    \langle W_2 A_{2,1}^{(a)} \rangle_0^c &=
    -\frac{N}{48V_D^2}\int_{xyu}
    \langle\tr\left(\left[\xi(x),\partial_0\xi(x)\right]
      \left[\xi(y),\partial_0\xi(y)\right]\right) 
    \xi_a(u)\xi_a(u)\rangle_0^c
    \\
    &=\frac{4N^2\Ns}{3V_D^2}\int_{xyu}
    G(x-u)G(y-u)\partial_0^x \partial_0^y G(x-y)
    \\
    &
    =\frac43 N^2\Ns\frac{1}{V_D}\IDR_{31}\,.
    \label{W2A21aav}
  \end{aligned}
\end{equation}

Furthermore
\begin{align}
  \langle W_2 A_{2,1}^{(b)} \rangle_0^c &=
  -\frac{1}{768V_D}\int_{x y u}
  \langle \tr\left(\left[\xi(x),\partial_0\xi(x)\right]
    \left[\xi(y),\partial_0\xi(y)\right]\right)
  \tr\left(\left[\xi(u),\partial_\mu\xi(u)\right]^2\right)\rangle_0^c
  \nonumber
  \\
  &=w_2^{(a)}+w_2^{(b)}+w_2^{(c)}\,,
\end{align}
with
\begin{equation}
  \begin{aligned}
    w_2^{(a)}&=\frac{N}{96V_D}G(0)\int_{x y u}
    \langle\tr\left(\left[\xi(x),\partial_0\xi(x)\right]
      \left[\xi(y),\partial_0\xi(y)\right]
      \partial_\mu\xi^a(u)\partial_\mu\xi^a(u)\right)\rangle_0^c
    \\
    &=-\frac{N}{12V_D}G(0)Z_1\int_{x y u}
    \partial_\mu^u G(x-u)\partial_\mu^u G(y-u)\partial_0^x\partial_0^{y} G(x-y)
    \\
    &=-\frac23 N^2\Ns\IDR_{10}\IDR_{21}\,.
  \end{aligned}
\end{equation}

\begin{equation}
  \begin{aligned}
    w_2^{(b)}&=\frac{N}{96V_D}\Box G(0)\int_{x y u}
    \langle\tr\left(\left[\xi(x),\partial_0\xi(x)\right]
      \left[\xi(y),\partial_0\xi(y)\right]
      \xi^a(u)\xi^a(u)\right)\rangle_0^c
    \\
    &=\frac23 N^2\Ns\frac{1}{V_D}\IDR_{31}\,. 
  \end{aligned}
\end{equation}
Note that $\IDR_{00}= -\Box G(0)= - 1/V_D$ since with dimensional
regularization one sets $\delta(0)=0$.
Finally
\begin{equation}
\begin{aligned}
    w_2^{(c)}&=-\frac{1}{96V_D}(Z_6+Z_7)\int_{x y u}
    G(x-u)\partial_0^{x2}G(x-u)
    G(y-u)\partial_0^{y2}G(y-u)
    \\
    &=-N^2\Ns\IDR_{21}^2\,.
\end{aligned}
\end{equation}

\subsection{Contribution from the 4-derivative terms}

For the averages we have
\begin{equation}\label{avC4i}
  \langle C_4^{(i)} \rangle_\mcA=
  \frac{1}{\Ns}\langle \left[C_4^{(i)}\right]\rangle_\mcA\,,
\end{equation}
where $\left[C_4^{(i)}\right]$ is obtained in Appendix~\ref{AppF} 
from $C_4^{(i)}$ by replacing
$\lambda^3_{ij}\lambda^3_{kl}$ by $\sum_a \lambda^a_{ij}\lambda^a_{kl}$
and averaging over the constant modes. From these expressions we obtain
\begin{equation}
  \begin{aligned}
    F_{3,1}&=-\frac{G_4^{(0)}}{N}\left\{\frac{1}{V_D}+2\Ns\IDR_{11}\right\}
    +G_4^{(1)}\left\{\frac{\Ns}{V_D}-2\IDR_{11}\right\}
    \\
    &+G_4^{(2)}\left\{\frac{1}{V_D}-N^2\IDR_{11}\right\}
    +\frac{G_4^{(3)}}{N}\left\{\frac{\Ns}{V_D}-(N^2-2)\IDR_{11}\right\}\,.
  \end{aligned}
\end{equation}
One can check that $F_{3,1}=0$ for $N=3$
when one sets $G_4^{(0)}=2\,,G_4^{(1)}=-1\,,G_4^{(2)}=-2\,,G_4^{(3)}=4$,
as required by~\eqref{mcLSU3id}.

Finally
\begin{equation}
  F_{4,1}=F_{5,1}=0\,.
\end{equation}

\subsection{Summary}
\label{summary_DR}

Collecting the results together, the
expansion of the susceptibility with DR is given by
\begin{equation}
  \chi = \frac{1}{2 g_0^2} \left( 1 + g_0^2 R_1
    + g_0^4 R_2 + \ldots \right)\,,
  \label{chiDR}
\end{equation}
with
\begin{equation}
  R_1 = -2N\IDR_{21}\,,
  \label{R1f}
\end{equation}
and
\begin{equation}
  R_2=R_2^{(a)}+R_2^{(b)}\,,
\end{equation}
with
\begin{align}
  R_2^{(a)} &= N^2\left\{ -\Wov
    +2\IDR_{21}\left[\IDR_{10}-\IDR_{21}\right]
    +\frac{4}{V_D}\IDR_{31} \right\}\,,
  \label{R2af}
  \\
  R_2^{(b)} &=-4\left[
-\frac{G_4^{(0)}}{N}\left\{\frac{1}{V_D}+2\Ns\IDR_{11}\right\}
+G_4^{(1)}\left\{\frac{\Ns}{V_D}-2\IDR_{11}\right\}\right.
\nonumber\\
&\left.+G_4^{(2)}\left\{\frac{1}{V_D}-N^2\IDR_{11}\right\}
+\frac{G_4^{(3)}}{N}\left\{\frac{\Ns}{V_D}-(N^2-2)\IDR_{11}\right\}\right]
  \label{R2bf}
  \\
  &=-\frac{4}{N}\left[-G_4^{(0)}+ N \Ns G_4^{(1)}+NG_4^{(2)}
    + \Ns G_4^{(3)}\right]\frac{1}{V_D}
  \nonumber\\
  &+\frac{4}{N}\left[2\Ns G_4^{(0)}+2N G_4^{(1)}+N^3G_4^{(2)}
    +(N^2-2)G_4^{(3)}\right]\IDR_{11}\,.
  \label{R2bfp}
\end{align}
For $N=2,3$ the relations~\eqref{GtransSU2},~\eqref{GtransSU3}
are satisfied.

\subsection{\boldmath Renormalization of the free energy in $d=4$}

We first recall some results obtained in~\cite{Niedermayer:2016ilf} 
for the behavior of the functions as $q\to0$:
\begin{align}
  \IDR_{10}&= -\beta_1(\ell)\Ls^{-2}+\order{q}\,,
  \\
  \IDR_{11}&=\frac{1}{\Ls^4}\left\{
    \frac12\left(1-q\ln\Ls\right)\left[\gamma_1(\ell)-\frac12\right]
    +q\mcW_1(\ell,\ellhat)\right\}+\order{q^2}\,,
  \label{I11}
  \\
  \IDR_{21}&=\frac{1}{8\pi\Ls^2}\left(\gamma_2(\ell)-1\right)
  + \order{q}\,,
  \label{I21}
  \\
  \IDR_{31}&= -\frac{1}{32\pi^2}\left[\frac{1}{q} - \ln\Ls
    -\frac12\gamma_3(\ell)\right]+\order{q}\,,
\end{align}
where the shape functions $\beta_1(\ell)$, $\gamma_i(\ell)$ and
$\mcWov(\ell)$ are given in~\cite{Niedermayer:2016yll},
and for the 2-loop function
\begin{equation}
  \Wov=\frac{1}{16\pi^2\Ls^4}
  \left\{\left[\frac{1}{q}-2\ln\Ls\right]\mcW_0(\ell)
    +\frac{1}{3\ell}\ln(\ellhat)
    -\frac{10}{3}\mcW_1(\ell,\ellhat)
    +\mcWov(\ell)\right\}+\order{q}\,,
  \label{Wov_d4}
\end{equation}
with the non-singular shape function~\cite{Niedermayer:2016yll}:
\begin{equation}
  \mcW_0(\ell)=\frac53\left(\frac12-\gamma_1(\ell)\right)
  -\frac{1}{3\ell}\,.
\end{equation}
The shape function $\mcW_1(\ell,\ellhat)$ occurring in~\eqref{I11} 
and~\eqref{Wov_d4} is not needed here (see below).

Below we switch to the conventional couplings $L_i=-G_4^{(i)}/4$ and 
express the bare couplings through the renormalized ones by
\begin{equation}
  L_i=\Lr{i}+ 
  \frac{v_i}{16\pi^2}\mu^{D-4} \left(\frac{1}{D-4}+\Cov \right) \,.
  \label{L_ren}
\end{equation}
where 
\begin{equation}
  \Cov=\log\cov=-\frac12\left( \ln(4\pi)-\gamma_E+1\right)=-1.476904292\,.
  \label{Cov}
\end{equation}
By convention~\cite{Gasser:1984gg}
the renormalized couplings are taken at the scale $\mu=M_\pi$,
where $M_\pi$ is the mass of the charged pion.

Requiring cancellation of the $\propto 1/(D-4)$ terms in $R_2$ one obtains
two relations,
\begin{equation}
  \begin{split} 
    &  Nv_2 - v_0 + (N^2-1)(Nv_1 + v_3) = \frac{5}{48}N^3 \,,
    \\
    & 2Nv_0 - Nv_3 + (N^2-2)(v_2-2v_1) = 0 \,.
  \end{split}
  \label{ueqs}
\end{equation}
Due to these relations the terms $\ln(\ellhat)$ and $\mcW_1(\ell,\ellhat)$
depending on auxiliary, unphysical box size, also cancel.
The relations~\eqref{ueqs} are satisfied by the
coefficients $v_i$ which were calculated in an elegant way 
by Gasser and Leutwyler~\cite{Gasser:1984gg}\footnote{The 
coefficients in~\cite{Gasser:1984gg} are written out explicitly only 
for $N=3$ but the previous steps are done for general $N$.
The $N=3$ coefficients $\Gamma_i$ in~\cite{Gasser:1984gg} are
given by $\Gamma_1=v_1+v_0/2=3/32$, $\Gamma_2=v_2+v_0=3/16$
and $\Gamma_3=v_3-2v_0=0$.}:
\begin{equation}
  v_0 = N/48 \,, \quad v_1 = 1/16 \,, \quad v_2 = 1/8  \,,  \quad v_3 = N/24 \,.
  \label{v_sol}
\end{equation}

Finally one has
\begin{equation} \label{chires}
 \Ls^2 \chi = \frac12 F^2 \Ls^2 \left( 1 + \frac{1}{F^2 \Ls^2} (\Ls^2 R_1)
    + \frac{1}{F^4\Ls^4} (\Ls^4 R_2) +\order{(F \Ls)^{-6}}
  \right)
\end{equation}
where
\begin{equation}
  \begin{split}
    \Ls^2 R_1 & = -\frac{N}{4\pi}(\gamma_2-1)
    \,,
    \\
    \Ls^4 R_2 & = -\frac{N^2}{32\pi^2}\left[(\gamma_2-1)^2 
      + 8\pi (\gamma_2-1)\beta_1 + 2 \mcWov -\frac{2}{\ell} \gamma_3 \right]
    \\
    & 
    + \frac{5N^2}{48\pi^2}\left[\frac{1}{\ell} - \gamma_1 + \frac12 \right]
    \log\left(\cov \Ls M_\pi\right) 
    \\
    & 
    -\frac{8}{N}\left[
      2(N^2-1)\Lr{0} + 2N\Lr{1} + N^3\Lr{2} + (N^2-2)\Lr{3}
    \right] \left(\gamma_1-\frac12\right)
    \\
    & 
    +\frac{16}{N}\left[
      -\Lr{0} + N(N^2-1)\Lr{1} + N\Lr{2} + (N^2-1)\Lr{3}
    \right] \frac{1}{\ell} \,,   \qquad (N\ge 4)\,.
  \end{split}
  \label{R2res}
\end{equation}

For $N=3$ one should here omit the term proportional to $\Lr{0}$. 
Similarly, for $N=2$ one should omit $\Lr{0}$ and $\Lr{3}$. 
In addition, to use the conventional notation (stemming from the O(4)
formulation), one should make the replacement
$\Lr{1}\to \lr{1}/4$, $\Lr{2}\to \lr{2}/4$.
This result is also invariant under the transformations corresponding
to~\eqref{GtransSU2} and~\eqref{GtransSU3}.

For the O($n$) case one has~\cite{Niedermayer:2016yll}
\begin{equation}
  \begin{split}
    L^4 R_2^{\mathrm{O}(n)} & = -\frac{n-2}{16\pi^2}
    \left[\left(\gamma_2-1\right)^2 +8\pi\left(\gamma_2-1\right)\beta_1
      + 2 \mcWov-\frac{(n-2)}{\ell}\gamma_3
    \right]
  \\
  & 
  + \frac{n-2}{24\pi^2}\left[\frac{3n-7}{\ell} 
    - 5 \left(\gamma_1 - \frac12\right) \right]
  \log\left(\cov L M_\pi\right) 
  \\
  & 
  -2  \left(2\lr{1}+n\lr{2}\right)\left(\gamma_1-\frac12\right)
  +4 \left((n-1) \lr{1}+\lr{2}\right) \frac{1}{\ell}\,.
  \end{split}
  \label{R2_On_res}
\end{equation}
Our result~\eqref{R2res} for $N=2$ flavors agrees with this taken at $n=4$.

\section{Computation of mass gap on a periodic strip}
\label{massgap}

In this section we will compute the mass gap of the 4d chiral
$\SSUN$ model on a periodic strip.
We will follow the method first used in~\cite{Luscher:1991wu} and later
in~\cite{Niedermayer:2010mx}. 
In the latter references the computation was done using
lattice regularization. Here we will employ dimensional regularization
as we did in~\cite{Niedermayer:2016yll}. The dynamical fields 
$U(x)$ are now defined in a volume
\begin{equation}
\Lambda=\left\{x;x_0\in[-T,T]\,,
x_\mu\in[0,L]\,,{\rm for}\,\mu=1,\ldots,d-1\,,
x_\mu\in[0,\Lhat]\,,{\rm for}\,\mu=d,\ldots,D-1\right\}\,,
\end{equation}
with periodic boundary conditions in the $D-1$ ``spatial'' directions,
and free boundary conditions in the time direction.

Here we will only give a brief description of the computation
since it follows closely that for the O($n$) model~\cite{Niedermayer:2016yll}.
We first compute the 2-point function
\begin{equation}
  \begin{aligned}
    C_0(x)&=\lim_{T\to\infty}\frac{1}{N}
    \langle\tr\left(U^\dagger(x)U(0)\right)\rangle
    \\
    &\propto\mre^{-(\mathcal{E}_1-\mathcal{E}_0)|x_0|}\,,\,\,\,\,
    (|x_0|\to\infty)\,.
  \end{aligned}
\end{equation}
It follows that the mass gap
\begin{equation}
  \mathcal{E}_1-\mathcal{E}_0=-\lim_{x_0\to\infty}\frac{\partial}{\partial x_0}
  \ln C(x_0)\,.
  \label{dc0d}
\end{equation}
Since $C_0(x)$ has a perturbative expansion of the form
\begin{equation}
  C_0(x)=1+\sum_{\nu=1}^\infty g_0^{2\nu} C_0^{(\nu)}(x)\,,
\end{equation}
equation (\ref{dc0d}) yields the power series 
\begin{equation}
  \mathcal{E}_1-\mathcal{E}_0=\frac{1}{2\VdD}\sum_{\nu=1}^\infty 
  g_0^{2\nu}\triangle^{(\nu)}\,.
\end{equation}
If for $x_0\to\infty$:
\begin{equation}
  C_0^{(\nu)}(x)\sim\sum_{r=0}^\nu\bar{c}_r^{(\nu)}
  \left(\frac{x_0}{2\VdD}\right)^r+{\rm exponentially\,\,damped}
\end{equation}
then
\begin{align}
  \triangle^{(1)}&=-\bar{c}_1^{(1)}\,,
  \\
  \triangle^{(2)}&=-\bar{c}_1^{(2)}+\bar{c}_1^{(1)}\bar{c}_0^{(1)}
  =-\bar{c}_1^{(2)}-\triangle^{(1)}\bar{c}_0^{(1)}\,,
  \\
  \triangle^{(3)}&=-\bar{c}_1^{(3)}+\bar{c}_1^{(2)}\bar{c}_0^{(1)}
  +\bar{c}_0^{(2)}\bar{c}_1^{(1)}-\bar{c}_1^{(1)}\bar{c}_0^{(1)2}
  \nonumber\\
  &=-\bar{c}_1^{(3)}-\triangle^{(2)}\bar{c}_0^{(1)}
  -\triangle^{(1)}\bar{c}_0^{(2)}\,.
  \label{triangle3}
\end{align}
It thus suffices to compute the coefficients $c_i^{(r)}$ with
$i=0,1$\footnote{Computation of higher coefficients $c_i^{(r)}\,i>1$
can serve as useful checks since these are fixed by the
requirement of exponentiation}.

The fields $U(x)$ are parameterized as in~\eqref{Uparam} but
now the $\xi(x)$-field satisfies Neumann boundary 
conditions~\cite{Luscher:1991wu}
\begin{equation} 
  \partial_0\xi(x)=0\,\,\,\,\,\,{\rm for}\,\,x_0=\pm T\,, 
\end{equation}
and periodic boundary conditions in the spatial directions.

The corresponding free 2-point function is given by
\begin{equation}
  \langle\xi_a(x)\xi_b(y)\rangle_0=\delta_{ab}G(x,y)\,,
\end{equation}
with 
\begin{equation}
  \begin{aligned}
    G(x,y)&=\frac{1}{\VdD}\left(
      \frac{x_0^2+y_0^2}{4T}-\frac12 |x_0-y_0|+\frac{T}{6}\right)
    \\
    &+\sum_{m=-\infty}^{\infty}\Bigl\{R\left(x_0-y_0+4mT,\bfx-\bfy\right)
    \\
    &+R\left(x_0+y_0+2(2m+1)T,\bfx-\bfy\right)\Bigr\}\,,
  \end{aligned}
  \label{Grepd}
\end{equation}
where
\begin{equation}
  R(z)=\frac{1}{2\VdD}\sum_{\bfp\ne0}\frac{1}{\omega_\bfp}
  \mre^{-\omega_\bfp|z_0|}\mre^{i\bfp\bfz}\,,
\end{equation}
where the sum goes over 
$p_\mu=\frac{2\pi\nu_\mu}{L_\mu}\,,\,\,\mu=1,\ldots,D-1$
with $\nu_\mu\in \Z$, and
\begin{equation}
  \omega_\bfp=|\bfp|\,.
\end{equation}

Expanding
\begin{equation}
  \frac{1}{N}\tr\left(U^\dagger(x)U(0)\right)
  =1+\sum_{\nu=1}^\infty g_0^{2\nu}\theta_\nu+\sum_{\nu=1}^\infty g_0^{2\nu+1}
  \rho_\nu\,,
\end{equation}
the operators $\rho_\nu$ are not of interest to us here since their
expectation values with operators even in $\xi$ are zero, and
the 2-point correlation function has a perturbative expansion of the form 
\begin{equation}
  \langle \frac{1}{N}\tr\left(U^\dagger(x)U(0)\right)\rangle
  =1+\sum_{\nu=1}^{\infty}g_0^{2\nu}\omega_\nu\,,
\end{equation}
with
\begin{align}
  \omega_1&=\langle\theta_1\rangle_0\,,
  \\
  \omega_2&=\langle\theta_2\rangle_0
  -\langle\theta_1\mcA_1\rangle_0^{\rm c}\,,
  \\
  \omega_3&=\langle\theta_3\rangle_0
  -\langle\theta_2 \mcA_1\rangle_0^{\rm c}
  -\langle\theta_1 \mcA_2\rangle_0^{\rm c}
  +\frac12\langle\theta_1 \mcA_1^2\rangle_0^{\rm c}\,,
\end{align}
where $\langle\dots\rangle^{\rm c}$ denote connected parts.

The interaction terms in the total action have the same form as 
in the previous section apart from the integration range which is now
$\Lambda$, and the volume factors $V_D$ in 
the expressions for $A_{2,1}^{(a)},A_{2,2}^{(a)},A_{2,2}^{(b)}$
should be replaced by $|\Lambda|=2T\VdD\,,\,\,\,\VdD=L^{d-1}\Lhat^{D-d}$.

The computation now proceeds as in~\cite{Niedermayer:2016yll},
and here we only give the final results. In lowest order
\begin{align}
\bar{c}_1^{(1)}&=-\frac{2\Ns}{N}\,,
\\
\bar{c}_0^{(1)}&=-\left(\frac{2\Ns}{N}\right)R(0)\,,
\end{align}
where $R(0)$ is dimensionally regularized~\cite{Niedermayer:2016yll}. 
So for the energy shift, first computed by Leutwyler~\cite{Leutwyler:1987ak},
\begin{equation}
\triangle^{(1)}=\frac{2\Ns}{N}\,.
\end{equation}

In the next order
\begin{align}
  \bar{c}_1^{(2)}&=\frac{2\Ns(\Ns-1)}{N^2}R(0)\,,
  \\
  \bar{c}_0^{(2)}&=\frac{\Ns(\Ns-1)}{N^2}R(0)^2\,,
\end{align}
yielding
\begin{equation} 
  \triangle^{(2)}=2\Ns R(0)\,.
\end{equation}

Finally at third order we obtain
\begin{equation}\label{barc13}
  \bar{c}_1^{(3)}=\frac{\Ns}{N^3}\left[N^4+2N^2-4\right]R(0)^2    
  -2\Ns N\left(W+\frac38 Y\right)+\bar{c}_1^{(3;13)}\,,
\end{equation} 
where
\begin{equation}
  W= -\int_{-\infty}^\infty\mrd z_0\,
  \int_{\mathbf{z}} R(z)^2\partial_0^2 R(z)\,,
\end{equation}
and
\begin{equation}
  Y=\frac{1}{\VdD^2}\sum_{\bfp\ne0}\frac{1}{\bfp^2}\,.
\end{equation}
The term $\bar{c}_1^{(3;13)}$ appearing in~\eqref{barc13}
is the contribution to the correlator from the 4-derivative terms:
\begin{equation}
  \bar{c}_1^{(3;13)}=-\frac{8\Ns}{N^2}
  \left[2(N^2-1)G_4^{(0)}+2NG_4^{(1)}+N^3G_4^{(2)}+(N^2-2)G_4^{(3)}\right]
  \ddot{R}(0)\,.
\end{equation}

The 3rd order energy shift is given by:
\begin{align}
  \triangle^{(3)}&=\Ns N\left[2W+\frac34 Y+R(0)^2\right]
  \nonumber\\
  &+\frac{8\Ns}{N^2}
  \left[2(N^2-1)G_4^{(0)}+2NG_4^{(1)}+N^3G_4^{(2)}+(N^2-2)G_4^{(3)}\right]
  \ddot{R}(0)\,.
\end{align}

Defining the moment of inertia $\Theta$ through\footnote{%
  For $N=2$ this is consistent with the standard definition of $\Theta$
  for O($4$).}  
\begin{equation} \label{m1_SUN}
  m_1=\frac{\Ns}{N\Theta}\,,
\end{equation}
then
\begin{equation}
  \frac{\Theta}{\VdD}=\frac{1}{g_0^2}\left[1+\Theta_1 g_0^2+\Theta_2 g_0^4
    +\dots\right]\,,
\end{equation}
with
\begin{align}
  \Theta_1&=-NR(0)\,,
  \\
  \Theta_2&=-\frac{N}{2\Ns}\triangle^{(3)}+N^2 R(0)^2
  \\
  &=-N^2\left(W+\frac38 Y-\frac12 R(0)^2\right)
  \nonumber\\
  &-\frac{4}{N}
  \left[2(N^2-1)G_4^{(0)}+2NG_4^{(1)}+N^3G_4^{(2)}+(N^2-2)G_4^{(3)}\right]
  \ddot{R}(0)\,.
  \label{Theta2}
\end{align}

For O($n$) we had~\cite{Niedermayer:2016yll} for $d=4$:
\begin{equation} \label{m1_On}
  m_1=\frac{(n-1)}{2\overline{\Theta}}\,,
\end{equation}
with
\begin{equation}
  \frac{\overline{\Theta}}{F^2L^3}=1+
  \overline{\Theta}_1(FL)^{-2}
 +\overline{\Theta}_2(FL)^{-4}+\dots
\end{equation}
and
\begin{align}
  \overline{\Theta}_1&=-(n-2)L^2R(0)\,,
  \\
  \overline{\Theta}_2&=(n-2)L^4\left[
    -2W+R(0)^2-\frac34 Y\right]
    +4\left(2l_1+nl_2\right)\ddot{R}(0)\,.
\end{align}
We can check (for $d=4$) using~\eqref{GLid} (and recalling
$l_1=-g_4^{(2)}/4\,,l_2=-g_4^{(3)}/4$) and setting $F^2=1/g_0^2$ that
\begin{align}
L^2\left[\Theta_1\right]_{N=2}&=\left[\overline{\Theta}_1\right]_{n=4}\,,
\\
L^4\left[\Theta_2\right]_{N=2}&=\left[\overline{\Theta}_2\right]_{n=4}\,.
\end{align}

\subsection{\boldmath Renormalization of the mass gap in $d=4$}

The mass gap does not lead to a new renormalization condition
beyond~\eqref{ueqs} required by the free energy considered in this paper.
As discussed in~\cite{Niedermayer:2017uyr},
the reason is that they are closely related: 
knowing $\Theta$ determines the mass spectrum of Hamiltonian states
and these determine the free energy.

In~\cite{Niedermayer:2016ilf} we find
\begin{equation}
  W = \frac{5}{24\pi^2}\ddot{R}(0)\left[\frac{1}{D-4}-\ln L \right]
  + \frac{c_w}{L^4} \,,
\end{equation}
with
\begin{equation}
  c_w=0.0986829798\,.
\end{equation}
Further
\begin{equation}
    -L^2 R(0)  = -L \IDR_{10}^{(3)} = \beta_1^{(3)} = 0.2257849594\,,
  \label{R0}
\end{equation}
and $L^4 \ddot{R}(0)$ can be expressed through the known shape 
coefficients as\footnote{The expression in the square brackets above 
converges exponentially fast for $\ell\to\infty$, already at $\ell=4$ 
it agrees to 9 digits with the limiting value.}
\begin{equation}
  \begin{split} \label{ddR0}
    -L^4 \ddot{R}(0)\equiv\rho &= 8\pi^2 \beta_2^{(3)}(1) 
    = \frac12 \alpha_{2}^{(3)}(1)+\frac34 
    = \lim_{\ell\to\infty} \left[ \frac12\left(\gamma_1^{(4)}(\ell)-\frac12\right)
      +\frac{1}{\ell} \right]
    \\
    &=0.8375369107 \,. 
  \end{split}
\end{equation}
Also
\begin{equation}
  Y = L^{-3} \IDR_{10}^{(3)} 
  = L^{-3} \left[ \overline{G}_1 \right]_\mathrm{HL}^{(d=3)}
  = -\beta_1^{(3)} L^{-4}\,.
\end{equation}

After introducing the renormalized couplings~\eqref{L_ren}
one obtains
\begin{equation}
  \begin{aligned}
    L^2 \Theta_1  & = N \beta_1^{(3)} \,, \\
    L^4 \Theta_2 & =\frac12 N^2 
    \left[\beta_1^{(3)}\left(\beta_1^{(3)} + \frac34\right)- 2 c_w\right]
    - \frac{5N^2\rho}{24\pi^2} \log(\cov L M)
    \\
    & \quad
    - \frac{16\rho}{N}
    \left[ 2(N^2-1) \Lr{0} + 2N\Lr{1} + N^3\Lr{2} + (N^2-2)\Lr{3}\right]\,,
    \qquad (N\ge 4) \,.
  \end{aligned}
  \label{T2res}
\end{equation}
Note that the combination of the renormalized couplings is the same as 
one of the combinations appearing in~\eqref{R2bfp}.
Again for $N=3$ one should omit $\Lr{0}$, while for $N=2$ the couplings 
$\Lr{0}$ and $\Lr{3}$ should be omitted, and $\Lr{1}$,  $\Lr{2}$ 
should be replaced by $4\lr{1}$, and $4\lr{2}$ respectively
(see Appendix~\ref{AppF}).

Comparing $\Theta_2$ with $R_2$ in~\eqref{R2res}, using the large-$\ell$
behavior of the shape coefficients from~\cite{Niedermayer:2017uyr} 
one finds that
\begin{equation} \label{limR1}
    \lim_{\ell\to\infty} \left(R_1+\frac{N}{6}\ell\right) = \Theta_1 \,, 
\end{equation}
\begin{equation} \label{limR2}
  \lim_{\ell\to\infty} R_2  = \Theta_2 \,.
\end{equation}
For the susceptibility calculated in $\chi$PT for the long cylinder geometry
this gives a remarkably simple result,
\begin{equation} \label{chi_tube}
  L_s^2 \chi = \frac{\Theta}{2L_s} -\frac{N}{12}\frac{L_t}{L_s} 
  + \order{\frac{1}{F^4 L_s^4}} \,.
\end{equation}

In the $\mathrm{O}(n)$ model for $\ell\to\infty$
one obtains $L_s^4 R_2 = \text{const} (n-2)(n-4) \ell^2 + \order{1}$,
in contrast to the $\SSUN$ model. It is interesting to observe
that in the cases of $n=2$ and $n=4$ the manifold $S_{n-1}$
on which the system is moving is a group manifold,
U(1) and $\SUtw$ with symmetries U(1)$\times$U(1) and
$\SUtw$, correspondingly. 
While for general O($n$) the expansion parameter for large $\ell$
is $\ell/(F^2 L_s^2)$, in these special cases the expansion parameter
seems to be $1/(F^2 L_s^2)$ (see eq.~(3.6) of ref.~\cite{Niedermayer:2017uyr}).

Eq.~\eqref{chi_tube} is obtained assuming $L_s \ll L_t \ll F^2 L_s^3$.
This is a high-temperature expansion for the spatially constant modes
and at the same time a low-temperature expansion for the 
$\mathbf{p}\ne 0$ modes.
The leading term, $\Theta/2L_s$ is the classical result.
The second one is the leading quantum correction; 
it appears both for $\On$ and for $\SSUN$, 
and does not depend on the dynamics.
Note that $ L_s^2 \chi \propto \langle T_3^2 \rangle =
 \langle C_2 \rangle/ (N^2-1)$ where $C_2$ is the quadratic Casimir invariant,
hence in the more natural choice $(N^2-1) L_s^2 \chi $ the curvature 
of the $\SUN$ manifold, $N(N^2-1)/12$ appears.

A specific feature of the $\SSUN$ case is that in 
the $\chi$PT result~\eqref{chi_tube}
the $\propto L_s/\Theta \sim 1/(F^{2}L_s^{2})$ 
term is absent,
i.e.\ the LEC's to NNL order are hidden in the first term alone.
Related to this observation,
there is a strong evidence that in the $\SSUN$ rotator approximation
(describing the contribution of the spatially constant modes)
there are no power-like corrections to the first two terms 
in~\eqref{chi_tube} for general $N$ (cf.~\cite{NiedermayerNew}).
For the $\SSUtw\simeq\Of$ case this can be shown 
analytically; writing
\begin{equation} \label{logz0_O4}
  \log(z_0(u)) = \log(\sqrt{\pi}/4) - \frac32 \log u + u + \phi(u) \,,
\end{equation}
from (A.38) of~\cite{Niedermayer:2017uyr} it follows that
the correction term $\phi(u)$ decreases faster than any power of $u$.
In fact it is extremely small already at $u=0.1$; 
one has $\phi(0.1)=-5.4\times 10^{-41}$.
For $N=3,4$ and $5$ this was shown numerically~\cite{NiedermayerNew}.

A derivation of the susceptibility from a $\SSUN$ rotator (for
general $N$) will be presented in a separate paper~\cite{NiedermayerNew}.
Suffice it here to say that in this scenario we have numerically 
shown absence of power-like corrections for $N=3,4$ and $5$.

The absence of a $\sim 1/\Theta^2$ term
in the $\SSUth$ rotator approximation   
does however not necessarily mean that a term
$\order{F^{-4} L_s^{-4}}$ cannot be present in~\eqref{chi_tube},
since the simple rotator model requires modifications 
in order to match $\chi$PT at higher orders. 

In~\eqref{limR1} the limit $\ell\to\infty$ is reached exponentially
fast, while in~\eqref{limR2} apart from the exponentially small corrections
there are $\propto 1/\ell$ corrections as well. 
This gives for the deviation of the susceptibility $\chi_{\text{rot}}$
calculated for the \emph{standard} rotator\footnote{with the standard
Hamiltonian proportional to the quadratic Casimir invariant $C_2$.}
from the $\chi$PT result $\chi$ in the NNL order
\begin{equation} \label{chi-chirot}
  \begin{aligned}
   F^4 L_s^4 \frac{\chi-\chi_{\text{rot}}}{\chi} 
   &= \frac{16}{N \ell}
    \left[ (2N^2-3)(\Lr{0}+\Lr{3}) + N(N^2+1)(\Lr{1}+\Lr{2})\right]
    \\
    & \qquad
    + \frac{5N^2}{16\pi^2 \ell}
    \left[
      \log\left(\cov \Ls M_\pi\right)
    + \frac12 \alpha_0^{(3)}(1)-\frac13 \right] + \ldots
    \,.
  \end{aligned}
\end{equation}
The omitted terms at this order are vanishing exponentially 
as $\ell\to\infty$. 
The $1/\ell$ term given above should come from the distortion of the
rotator spectrum in the region of energies $E \ll 1/L_s$, 
much below the threshold for the $\mathbf{p}\ne 0$ modes.
In other words, the true rotator Hamiltonian differs 
from that of the standard rotator in higher order.
A similar situation was found in~\cite{Niedermayer:2016yll}
for the case of the O($n$) model.
The corresponding correction for the $\SSUN$ case is
discussed in~\cite{NiedermayerNew}.

Finally we make a few remarks concerning the sensitivity
of the observables on the 4-derivative couplings $\Lr{i}$.
The sensitivity of the isospin susceptibility at $\ell=1$
(hypercubic lattice) is obtained from~\eqref{R2res}
(observing that $\gamma^{(4)}(1)=0$)
\begin{equation} \label{sens_cubic}
   \frac{\delta_L \chi}{\chi} = \frac{1}{F^4 L_s^4}
   \left( 136 \delta\!\Lr{1} + 52 \delta\!\Lr{2} + 52 \delta\!\Lr{3} \right)\,,
   \quad (N=3\,,\;\ell=1)\,. 
\end{equation}

For a long cubic tube, $\ell\gg 1$, the sensitivity
of the susceptibility and of the mass gap on $\Lr{i}$ are
\begin{equation} \label{sens_tube}
  \begin{aligned}
  \frac{\delta_L \chi}{\chi} &= \frac{\delta_L m_1}{m_1} = 
  -\frac{16}{3F^4 L_s^4} \rho
  \left( 6 \delta\!\Lr{1} +27 \delta\!\Lr{2} + 7 \delta\!\Lr{3} \right)\,,\\
  &= \frac{1}{F^4 L_s^4}\left( -26.8 \delta\!\Lr{1} - 120.6\delta\!\Lr{2}
    -31.3\delta\!\Lr{3} \right) \,,
  \quad (N=3\,,\;\ell\gg 1) \,.
  \end{aligned}
\end{equation}
Note that all coefficients change sign as $\ell$ varies from 1 to
$\infty$; this feature can be used to select optimal values of $\ell$
for certain purposes e.g.\ to reduce the influence on the uncertainty
of the $\Lr{i}$'s on determination of $F$.

\begin{appendix}

\section{\boldmath $\SUN$ Gell-Mann matrices}
\label{AppA}

The $N\times N$ Gell-Mann hermitian $\lambda-$matrices satisfy
\begin{align}
\tr \lambda^a&=0\,,
\\
\tr\left(\lambda^a\lambda^b\right)&=2\delta_{ab}\,,
\\
\lambda^a\lambda^b&=\frac{2}{N}\delta_{ab}
+\left(d_{abc}+if_{abc}\right)\lambda^c\,,
\end{align}
where $f_{abc}$ is totally anti-symmetric and $d_{abc}$ is totally
symmetric and
\begin{equation}
\sum_a d_{aac}=0\,.
\end{equation}

Note the identities
\begin{align}
&f_{abc}f_{cde}+f_{dbc}f_{ace}+f_{ebc}f_{adc}=0\,,
\\
&f_{abc}d_{cde}+f_{dbc}d_{ace}+f_{ebc}d_{adc}=0\,,
\\
&f_{abc}f_{dec}=\frac{2}{N}\left(\delta_{ad}\delta_{be}-\delta_{ae}\delta_{bd}
\right)+d_{adc}d_{bec}-d_{aec}d_{bdc}\,,
\end{align}
and
\begin{align}
f_{abc}f_{dbc}&=N\delta_{ad}\,,
\\
d_{abc}d_{dbc}&=\frac{(N^2-4)}{N}\delta_{ad}\,.
\end{align}

Completeness reads
\begin{equation}
\sum_a \lambda^a_{ij}\lambda^a_{kl}=2\delta_{il}\delta_{jk}
-\frac{2}{N}\delta_{ij}\delta_{kl}\,.
\end{equation}
From this we immediately get
\begin{equation}
\sum_a \lambda^a\lambda^a=\frac{2\Ns}{N}\one\,, \qquad
\sum_a \lambda^a\lambda^b\lambda^a=-\frac{2}{N}\lambda^b\,,
\end{equation}
and
\begin{align} \label{compl1}
\sum_a \tr\left(\lambda^a A\right)\tr\left(\lambda^a B\right)
& =2\tr(AB)-\frac{2}{N}\tr(A)\tr(B)\,,
\\
\sum_a \tr\left(\lambda^a A\lambda^a B\right)
& =2\tr(A)\tr(B)-\frac{2}{N}\tr(AB)\,.
\label{compl2}
\end{align}

For $N=2\,,\,\,\lambda^a=\sigma^a$, the Pauli matrices.
Also for an SU(2) matrix
\begin{equation}
U=\exp\left(i\sum_{a=1}^3v_a\sigma^a\right)
=\cos(\sqrt{v^2})+i\frac{\sin(\sqrt{v^2})}{\sqrt{v^2}}
\sum_{a=1}^3v_a\sigma^a\,.
\end{equation}

Note for $N=3$ we have the extra identity~\cite{MacFarlane:1968vc}
\begin{equation} \label{SU3id}
d_{abc}d_{cde}+d_{dbc}d_{ace}+d_{ebc}d_{adc}=\frac13\left(
\delta_{ab}\delta_{de}+\delta_{ad}\delta_{be}+\delta_{ae}\delta_{bd}\right)\,.
\end{equation}

\subsection{Group factors appearing in the perturbative computation} 

\begin{align}\label{Z1def}
  Z_1 &\equiv -\sum_{a,b}
  \tr\left(\left[\lambda^a,\lambda^b\right]
    \left[\lambda^a,\lambda^b\right]\right)
  = 8N\Ns\,,
  \\
  Z_2&\equiv\tr\left(\left[\lambda^a,\left[\lambda^b,\lambda^c\right]\right]
    \left[\lambda^a,\left[\lambda^b,\lambda^c\right]\right]\right)
  \notag
  \\
  &=32 f_{bce}f_{aeg}f_{bcd}f_{adg}= 32N^2\Ns\,,
  \\
  Z_3&\equiv\tr\left(\left[\lambda^a,\left[\lambda^b,\lambda^c\right]\right]
    \left[\lambda^b,\left[\lambda^a,\lambda^c\right]\right]\right)
  \notag
  \\
  &=32 f_{bce}f_{aeg}f_{bad}f_{cdg}= 16N^2\Ns\,,
  \\
  Z_4&\equiv\tr\left(\left[\lambda^a,\left[\lambda^a,\lambda^b\right]\right]
    \left[\lambda^c,\left[\lambda^c,\lambda^b\right]\right]\right)
  =32N^2\Ns\,.
\end{align}
\begin{equation}
  \begin{aligned}
    Z_5&\equiv\tr\left(\left[\lambda^a,\lambda^b\right]
      \left(
        \left[\lambda^c,\left[\lambda^c,\left[\lambda^a,\lambda^b
            \right]\right]\right]
       +\left[\lambda^c,\left[\lambda^a,\left[\lambda^c,\lambda^b
            \right]\right]\right]\right.\right.
    \\
    &+\left.\left.
        \left[\lambda^a,\left[\lambda^c,\left[\lambda^c,\lambda^b
            \right]\right]\right]\right)\right)
    =-Z_2-Z_3-Z_4=-80N^2\Ns\,.
  \end{aligned}
\end{equation}

\begin{align}
  Z_6&\equiv \tr\left(\left[\lambda^a,\lambda^b\right]
    \left[\lambda^c,\lambda^d\right]\right)
  \tr\left(\left[\lambda^a,\lambda^b\right]
    \left[\lambda^c,\lambda^d\right]\right)
  \nonumber
  \\
  &=64 f_{abe}f_{cde}f_{abg}f_{cdg}=64 N^2\Ns\,,
  \\
  Z_7&\equiv \tr\left(\left[\lambda^a,\lambda^b\right]
    \left[\lambda^c,\lambda^d\right]\right)
  \tr\left(\left[\lambda^a,\lambda^c\right]
    \left[\lambda^b,\lambda^d\right]\right)
  \nonumber
  \\
  &=64 f_{abe}f_{cde}f_{acg}f_{bdg}=32 N^2\Ns\,,
  \\
  \label{Z8def}
  Z_8&\equiv\sum_{a,b,c}\tr\left(\lambda^a\lambda^b\lambda^c
                   \lambda^a\lambda^b\lambda^c\right)
  =\frac{8\Ns(N^2+1)}{N^2}\,. 
\end{align}

\section{Proof of eq.~\eqref{mcLSU3id}}
\label{AppB}

We start from the trivial identity
\begin{equation}
\tr (ABAB) = \frac12 \tr(\{A,B\}^2)-\tr(A^2B^2)\,.
\end{equation}
Let $A,B$ be traceless SU($N$) matrices $A=A_a\lambda^a, B=B_a\lambda^a$;
we have
\begin{align}
\tr(A^2B^2)&=A^aA^bB^cB^d\left[\frac{4}{N}\delta_{ab}\delta_{cd}
+2d_{abe}d_{cde}\right]\,,
\\
\tr(ABAB)&=-\tr(A^2B^2)+2A^aB^bA^cB^d
\left[\frac{4}{N}\delta_{ab}\delta_{cd}
+2d_{abe}d_{cde}\right]\,.
\end{align}
So 
\begin{align}
&\tr(ABAB)+2\tr(A^2B^2)=2\left[2A^aB^bA^cB^d+A^aA^bB^cB^d\right]
\left[\frac{2}{N}\delta_{ab}\delta_{cd}+d_{abe}d_{cde}\right]
\\
&=2A^aA^bB^cB^d\left[
d_{abe}d_{cde}+d_{ace}d_{bde}+d_{ade}d_{bce}+\frac{2}{N}\left(
\delta_{ab}\delta_{cd}+\delta_{ac}\delta_{bd}+\delta_{ad}\delta_{bc}\right)
\right]\,.
\end{align}
For SU(3) we have using~\eqref{SU3id}
\begin{align} \label{SU3idx}
&\tr(ABAB)+2\tr(A^2B^2)=2A^aA^bB^cB^d\left(
\delta_{ab}\delta_{cd}+\delta_{ac}\delta_{bd}+\delta_{ad}\delta_{bc}\right)
\\
&=\frac12\tr(A^2)\tr(B^2)+\tr^2(AB)\,,\,\,\,\,N=3\,. 
\end{align}

Now consider $U(x)$ slowly varying in $x$. 
Define $U(x)=U(0)V(x)$; then close to $x=0$, $V(x)$ is close to 
the identity matrix
\begin{equation}
  V(x)=1+i x_\mu A_\mu +\ldots
\end{equation}
where $A_\mu$ are traceless hermitian matrices.
Then the $\mcL_i$ at $x=0$ can be computed by replacing
$\partial_\mu U$ by $iU(0)A_\mu$. The factors involving $U(0)$
cancel and then~\eqref{mcLSU3id} follows using~\eqref{SU3idx}.

\section{\boldmath Redundancy of 
  $\tr(\Box U^\dagger \Box U)$}
\label{AppC}

For the SU(2) case the 4-derivative operator
\begin{equation}
  \mcL_2^{(X)} = \tr\left( \Box U^\dagger \Box U \right)
  \label{L2X}
\end{equation}
corresponding to $(\Box S \cdot \Box S)$ in O(4),
turns out to be redundant: it can be eliminated by changing the 
integration variable $U(x)$ in the path integral.
Below we show that this remains true for general SU($N$). 

Consider the change of variables
\begin{equation}
  U \to U \mre^{\alpha F} = U \left( 1 + \alpha F + \order{\alpha^2}  \right)
\end{equation}
where $F$ is a traceless anti-hermitian matrix.

By choosing 
\begin{equation}
  F = \frac12 \left( U^\dagger \Box U - \Box U^\dagger U \right)\,,
  \label{Fdef}
\end{equation}
one has
\begin{equation}
  U \to  U + \alpha\, U F 
  = U + \frac{\alpha}{2}\left(\Box U - U \Box U^\dagger U\right)
  + \order{\alpha^2}\,.
\end{equation}

For the SU(2) case this corresponds to the change of variables
\begin{equation}
  S \to S + \alpha \left[ \Box S - S (S\cdot \Box S)\right]
  + \order{\alpha^2}\,,
\end{equation}
which is the transformation used to show the redundancy of the
operator $\tr(\Box S \, \Box S)$.

We have still to show that $F$ is indeed traceless. One has 
\begin{equation}
  \tr F  = \frac12 \tr\left(U^\dagger \Box U - \Box U^\dagger U \right) 
  = \mathrm{Im} \, \tr\left(U^\dagger \Box U\right)\,.
  \label{trF}
\end{equation}
Further we can write
\begin{equation}
  \left. \mathrm{Im} \,\tr\left( U^\dagger(x)\Box  U(x)\right)\right|_{x=0}
  = \left.\mathrm{Im} \, \tr\left[ \Box \left(U^\dagger(0) U(x)\right)
    \right]\right|_{x=0}\,.
\end{equation}
One has
\begin{equation}
W(x)~\equiv U^\dagger(0) U(x) = 
\exp\left( i x_\mu A_\mu +\frac12 i x_\mu x_\nu B_{\mu\nu}+\order{x^3}\right)\,,
\end{equation}
where $A_\mu$ and $B_{\mu\nu}$ are traceless hermitian matrices.
From this it follows that
\begin{equation}
\left. \mathrm{Im} \, \Box W(x)\right|_{x=0} = \mathrm{Im} \,\tr
\left(-A_\mu A_\mu + i B_{\mu\mu} \right) = 0\,.
\end{equation}
Therefore we can conclude that the operator $\tr(\Box U^\dagger\, \Box U)$
can be transformed away by a field redefinition.

A similar discussion to that presented above has been given 
by Leutwyler in eq.~(11.6) and the following paragraph
of his article~\cite{Leutwyler:1991mz}.

\section{Faddeev--Popov trick for the zero modes}
\label{AppD}

Consider the SU($N$) partition function formally given by
\begin{equation}
Z=\int\left[\prod_x\mrd U(x)\right]\mre^{-A(U)}\,.
\end{equation}
We parameterize $U$ as in~\eqref{Uparam}.
The integral over the constant $u$ factors out for this consideration.
The action and measure are invariant under global SU($N$) transformations
\begin{equation}
U(x)\to VU(x)\,,
\end{equation}
which induces a change 
\begin{equation}
\xi_a(x)\to  \xi_a^V(x)\,.
\end{equation}
Define a la Faddeev--Popov $\Phi[\xi]$ through the integral
\begin{equation} \label{Phidef}
1=\Phi[\xi]\int\mrd V
\prod_a^{N^2-1}\delta\left(\int_x \xi_a^V(x)\right)\,. 
\end{equation}
Now the action, measure and also 
$\Phi[\xi]$ are invariant under SU($N$) transformations,
so inserting 1 in the form of the rhs of~\eqref{Phidef}
in the partition function we obtain
\begin{equation}
Z=\int\mrd V\int\prod_x\left[\frac{\mrd \xi(x)}{M[\xi(x)]}\right]
\mre^{-A(\xi)}\Phi[\xi]\prod_{a=1}^{N^2-1}\delta\left(\int_x\xi_a(x)\right)\,.
\end{equation}
The group volume is an irrelevant factor. 
Also for DR we set $M[\xi(x)]=1\,,\,\,\forall x$.

Now we only need $\Phi[\xi]$ near the surface $\int_x\xi(x)=0$
and we can consider an infinitesimal transformation
\begin{equation}
V=1+i\alpha_a\lambda^a+\order{\alpha^2}\,. 
\end{equation}
This induces a change 
\begin{equation}
g_0\xi_a^V(x)\lambda^a=g_0\xi_a(x)\lambda^a+\alpha_a t^a(x)+\order{\alpha^2}\,,
\end{equation}
with $t(x)$ obtained by solving (the argument $x$ understood)
\begin{equation}
(1+i\alpha_a\lambda^a+\order{\alpha^2})\mre^{ig_0\xi_b\lambda^b}
=\mre^{ig_0\xi_a\lambda^a+i\alpha_a t^a}+\order{\alpha^2}\,,
\end{equation}
thereby yielding
\begin{align}
t^a &= \lambda^a + g_0\frac{i}{2}\left[\lambda^a,\xi\right] 
-g_0^2\frac{1}{12}
\left[\xi,\left[\xi,\lambda^a\right]\right]
\nonumber\\
&-g_0^4\frac{1}{720}
\left[\xi,\left[\xi,\left[\xi,
\left[\xi,\lambda^a\right]\right]\right]\right]+\dots
\\
&=T_{ab}\lambda^b\,,
\end{align}
with
\begin{align}
T_{ab}&=\frac12\tr\left(\lambda^b t^a\right)
\\
&=\delta_{ab}+g_0f_{abc}\xi_c
-g_0^2\frac{1}{24}\tr\left(\lambda^b\left[\xi,\left[\xi,\lambda^a\right]\right]
\right)
\nonumber\\
&-g_0^4\frac{1}{1440}
\tr\left(\lambda^b\left[\xi,\left[\xi,\left[\xi,
\left[\xi,\lambda^a\right]\right]\right]\right]\right)+\dots
\end{align}

So
\begin{align}
\Phi[\xi]^{-1}&=\int\prod_a
\left[\mrd\alpha_a\delta(\alpha_b\overline{T}_{ab}[\xi])\right]
\\
&=\left(\det\overline{T}[\xi]\right)^{-1}\,,
\end{align}  
with (setting $\int_x\xi_a(x)=0$)
\begin{equation}
\overline{T}_{ab}[\xi]=\delta_{ab}-g_0^2\overline{T}^{(1)}_{ab}
-g_0^4\overline{T}^{(2)}_{ab}+\dots
\end{equation}
where
\begin{align}
\overline{T}^{(1)}_{ab}&=\frac{1}{24V_D}\int_x
\tr\left(\lambda^b\left[\xi(x),\left[\xi(x),\lambda^a\right]\right]\right)\,,
\\
\overline{T}^{(2)}_{ab}&=\frac{1}{1440V_D}\int_x
\tr\left(\lambda^b\left[\xi(x),\left[\xi(x),\left[\xi(x),
\left[\xi(x),\lambda^a\right]\right]\right]\right]\right)\,.
\end{align}

The zero mode action is then given by
\begin{align}
A_{\rm zero} &=-\ln\Phi[\xi]
\\
&=-\tr\ln\left(\overline{T}[\xi]\right)
\\
&=-(N^2-1)\ln V_D+g_0^2\tr\overline{T}^{(1)}
+g_0^4\tr\left\{\overline{T}^{(2)}+\frac12\overline{T}^{(1)2}\right\}
+\dots
\end{align}
We have in particular 
\begin{equation}
\tr\overline{T}^{(1)}=\frac{N}{3V_D}\int_x\sum_a\xi_a(x)^2\,.
\end{equation} 

\section{\boldmath Some integrals over SU($N$) matrices}
\label{AppE}

Integrals with Haar measure over SU($N$) matrices $u$
(see e.g.~\cite{Creutz:1978ub}):  
\begin{equation}
\int\mrd u =1\,.
\end{equation}
\begin{equation} 
\int\mrd u\,u_{ij}(u^\dagger)_{kl} =\frac{1}{N}\delta_{il}\delta_{jk}\,.
\end{equation}
It follows
\begin{align} \label{intuudagger}
\int\mrd u\,\tr(uAu^\dagger B) &=\frac{1}{N}\tr(A)\tr(B)\,,
\\
\int\mrd u\,\tr(uA)\tr(u^\dagger B) &=\frac{1}{N}\tr(AB)\,.
\label{intuudagger2}
\end{align}

\begin{equation}
\int\mrd u\, u_{i_1j_1}\dots u_{i_N j_N}=\frac{1}{N!}
\epsilon_{i_1\dots i_N}\epsilon_{j_1\dots j_N}\,,
\end{equation}
where $\epsilon_{i_1\dots i_N}$ is the totally antisymmetric tensor
with $\epsilon_{1\dots N}=1\,$.

Note for $N=2$ and an SU(2) matrix $V$:
\begin{equation}
\epsilon_{ik}\epsilon_{jl}V_{kl}=V_{ij}^*=\left(V^\dagger\right)_{ji}\,,
\end{equation}
(the conjugate of the fundamental representation is equivalent to
the fundamental representation for SU(2)).
So
\begin{equation}
\int\mrd u\,\tr(uA)\tr(uB)=\frac12\tr(AB^\dagger)\,,\,\,\,\,N=2\,.
\end{equation}

\section{Expressions involving the 4-derivative terms}
\label{AppF}

The 4-derivative terms $B_4^{(i)}\,,C_4^{(i)}$ appearing
in~\eqref{A4h_DR} are given by:
\begin{align}
  B_4^{(0)}&=-i\int_x\tr\left(\left[\lambda^3, U^\dagger(x)\right]
    \partial_\mu U(x)\partial_0 U^\dagger(x) \partial_\mu U(x)
  \right)+\mathrm{h.c.}
  \\
  C_4^{(0)}&=\frac14\int_x\tr\left\{ 
    \partial_\mu U^\dagger(x)\left[\lambda^3,U(x)\right]
    \partial_\mu U^\dagger(x)\left[\lambda^3,U(x)\right]\right.
  \nonumber\\
  &+\partial_0 U(x)\partial_0 U^\dagger(x) 
  \left[\lambda^3,U(x)\right]\left[\lambda^3,U^\dagger(x)\right] 
  \nonumber\\
  &+\left.\partial_0 U^\dagger(x) \partial_0 U(x) 
    \left[\lambda^3,U^\dagger(x)\right]\left[\lambda^3,U(x)\right]
  \right\}+\mathrm{h.c.}\,.
\end{align}

\begin{align}
  B_4^{(1)}&=-i\int_x\tr\left( \partial_0 U^\dagger(x)
    \left[\lambda^3, U(x)\right]\right)
  \tr\left(\partial_\mu U^\dagger(x)\partial_\mu U(x) \right)
  +\mathrm{h.c.}\,,
  \\
  C_4^{(1)}&=\frac14\int_x\left\{
    \tr \left( \left[\lambda^3,U^\dagger(x)\right]
      \left[\lambda^3,U(x)\right] \right)
    \tr \left( \partial_\mu U^\dagger(x) \partial_\mu U(x) \right)\right.
  \nonumber\\
  &+\tr\left(\left[\lambda^3,U^\dagger(x)\right]\partial_0 U(x)\right)
  \tr\left(\left[\lambda^3,U^\dagger(x)\right]\partial_0 U(x)\right)
  \nonumber\\
  &+\left.\tr\left(\left[\lambda^3,U^\dagger(x)\right]\partial_0 U(x)\right)
    \tr\left(\partial_0 U^\dagger(x)\left[\lambda^3,U(x)\right]\right)
  \right\}
  +\mathrm{h.c.}\,.
\end{align}

\begin{align}
  B_4^{(2)}&=-i\int_x 
  \tr\left(\left[\lambda^3,U^\dagger(x)\right]\partial_\mu U(x)\right)
  \tr\left(\partial_0 U^\dagger(x)\partial_\mu U(x)\right)
  +\mathrm{h.c.}\,,
  \\
  C_4^{(2)}&=\frac14\int_x\left\{
    \tr\left(\left[\lambda^3,U^\dagger(x)\right]\left[\lambda^3,U(x)\right]\right)
    \tr\left(\partial_0 U^\dagger(x)\partial_0 U(x)\right)\right.
  \nonumber\\
  &+\tr\left(\left[\lambda^3,U^\dagger(x)\right]\partial_\mu U(x)\right)
  \tr\left(\left[\lambda^3,U^\dagger(x)\right]\partial_\mu U(x)\right)
  \nonumber\\
  &+\left.\tr\left( \left[\lambda^3,U^\dagger(x)\right]\partial_0 U(x)\right)
    \tr\left( \partial_0 U^\dagger(x)\left[\lambda^3,U(x)\right]\right)
  \right\}
  +\mathrm{h.c.}\,.
\end{align}

\begin{align}
  B_4^{(3)}&=-i\int_x\tr\left(\left[\lambda^3,U^\dagger(x)\right]\partial_0 U(x)
    \partial_\mu U^\dagger(x)\partial_\mu U(x)\right)
  +\mathrm{h.c.}\,,
  \\
  C_4^{(3)}&=\frac14\int_x\tr\left\{
    \left[\lambda^3,U^\dagger(x)\right]\left[\lambda^3,U(x)\right] 
    \partial_\mu U^\dagger(x) \partial_\mu U(x) \right.
  \nonumber\\
  &+  \left[\lambda^3,U^\dagger(x)\right]\partial_0 U(x)
  \left[\lambda^3,U^\dagger(x)\right]\partial_0 U(x) \
  \nonumber\\
  &+\left.\left[\lambda^3,U^\dagger(x)\right]\partial_0 U(x)
    \partial_0 U^\dagger(x)\left[\lambda^3,U(x)\right]\right\}
  +\mathrm{h.c.}\,.
\end{align}

Using~\eqref{compl1},~\eqref{compl2},~\eqref{intuudagger} 
and~\eqref{intuudagger2} the average 
$\left[C_4^{(0)}\right]$ in~\eqref{avC4i} is given by
\begin{align}
  \left[C_4^{(0)}\right]&=\frac14\int_x\int\mrd u\,\tr\left\{ 
    \partial_\mu \Uov^\dagger(x)u^\dagger\left[\lambda^a,u\Uov(x)\right]
    \partial_\mu \Uov^\dagger(x)u^\dagger\left[\lambda^a,u\Uov(x)\right]\right.
  \nonumber\\
  &+u\partial_0 \Uov(x)\partial_0 \Uov^\dagger(x)u^\dagger 
  \left[\lambda^a,u\Uov(x)\right]\left[\lambda^a,\Uov^\dagger(x)u^\dagger\right] 
  \nonumber\\
  &+\left.\partial_0 \Uov^\dagger(x)\partial_0 \Uov(x) 
    \left[\lambda^a,\Uov^\dagger(x)u^\dagger\right]\left[\lambda^a,u\Uov(x)\right]
  \right\}+\mathrm{h.c.}
  \\
  &=\frac12\int_x\int\mrd u\,\left\{
    2\tr\left(u\partial_\mu \Uov(x)\right)
    \tr\left(u^\dagger\partial_\mu \Uov^\dagger(x)\right)
  \right.
  \nonumber\\  
  &+\tr\left(u\Uov(x)\right)\left(
    \tr\left(u^\dagger\Uov^\dagger(x)\partial_0\Uov(x)
      \partial_0\Uov^\dagger(x)\right)
    +\tr\left(u^\dagger\partial_0\Uov^\dagger(x)
      \partial_0\Uov(x)\Uov^\dagger(x)\right)
  \right)
  \nonumber\\
  &+\tr\left(u^\dagger\Uov^\dagger(x)\right)\left(
    \tr\left(u\partial_0\Uov(x)\partial_0\Uov^\dagger(x)\Uov(x)\right)
    +\tr\left(u\Uov(x)\partial_0\Uov^\dagger(x)\partial_0\Uov(x)\right)
  \right)
  \nonumber\\
  &\left.-4N\tr\left(\partial_0\Uov(x)\partial_0\Uov^\dagger(x)\right)
  \right\}
  +\mathrm{h.c.}
  \\
  &=\frac{2}{N}\int_x\left\{
    \tr\left(\partial_\mu\Uov(x)\partial_\mu\Uov^\dagger(x)\right)
    -2\Ns\tr\left(\partial_0\Uov(x)\partial_0\Uov^\dagger(x)\right)
  \right\}\,.
\end{align}

Similarly for $\left[C_4^{(i)}\right]\,,i=1,2,3$ one obtains
\begin{align}
  \left[C_4^{(1)}\right]&=-2\int_x\,\left\{
    \Ns\tr\left( \partial_\mu \Uov(x)\partial_\mu \Uov^\dagger(x) \right)
    +2\tr\left(\partial_0\Uov(x)\partial_0\Uov^\dagger(x)\right)\right\}\,,
\\
  \left[C_4^{(2)}\right]&=-2\int_x\,\left\{
    N^2\tr\left(\partial_0 \Uov(x) \partial_0 \Uov^\dagger(x)\right)
    +\tr\left(\partial_\mu\Uov(x)\partial_\mu\Uov^\dagger(x)\right)\right\}\,,
\\
  \left[C_4^{(3)}\right]&=-\frac{2}{N}\int_x\,\left\{
    (N^2-2)\tr\left(\partial_0 \Uov(x)\partial_0 \Uov^\dagger(x)\right)
    +\Ns\tr\left(\partial_\mu\Uov(x)\partial_\mu\Uov^\dagger(x)\right)\right\}\,.
\end{align}

\section{\boldmath Some relations for the O($4$) 
  couplings}
\label{AppG}

Some relations between different conventions for the O(4) couplings 
to connect with those used in ref.~\cite{Niedermayer:2016yll}
\begin{equation}
  l_i=\lr{i} + \frac{w_i}{16\pi^2}\left(\frac{1}{D-4}+\log(\cov M)\right)
\end{equation}
where $w_1=n/2-5/3$, $w_2=2/3$ and choosing the scale $\mu=M$, 
the mass of the charged pion.

Further
\begin{equation}
  l_i=\frac{w_i}{16\pi^2}\left(\frac{1}{D-4}+\log(\cov \Lambda_i)\right)\,.
\end{equation}
From here
\begin{equation}
  \lr{i} =\frac{w_i}{16\pi^2} \log\left(\Lambda_i/M\right) 
    = \frac{w_i}{32\pi^2} \overline{l}_i
\end{equation}
since $\overline{l}_i = \log\left(\Lambda_i^2/M^2\right)$.

\end{appendix}

\end{document}